\documentclass[12pt, a4paper]{article}
 \usepackage{amsmath}
 \usepackage{amsfonts}
\begin{document}
\newcommand{\Dx}{\Delta q}
\newcommand{\ep}{\epsilon}
\newcommand{\qarr}{\stackrel{\cal Q}{\longrightarrow}}
\newcommand{\lc}{L^2 (\Sg_3 (t);\ \bf C;\ b^3 (t){\sqrt \omega (t)} d^3 q)}
\newcommand{\vp}{\varphi}
\newcommand{\1}{{\bf\hat 1}}
\newcommand{\Oc}{O\left(c^{-2(L+1)}\right)}
\newcommand{\im}{{\rm i}} 
\newcommand{\Sg}{\Sigma}
\newcommand{\bgt}{\bigotimes}
\newcommand{\ptl}{\partial}
\newcommand{\Sche}{ Schr\"odinger equation\ }
\newcommand{\eu}{$ E_{1,3} $}
\newcommand{\euf}{E_{1,3}}
\newcommand{\rif}{V_{1,3}}
\newcommand{\ri}{$V_{1,3}$ }
\newcommand{\ov}{\overline}
\newcommand{\stc}{\stackrel}
\newcommand{\defst}{\stackrel{def}{=}}
\newcommand{\qstc}{\stackrel{\cal Q}{\longrightarrow}}
\newcommand{\h}{\hbar}
\newcommand{\beq}{\begin{equation}}
\newcommand{\nde}{\end{equation}}
\newcommand{\beqa}{\begin{eqnarray}}
\newcommand{\ndea}{\end{eqnarray}}
\newcommand{\rin}{$V_{1,n}$}
\newcommand{\rinf}{V_{1,n}}
\newcommand{\rn}{$V_n$}
\newcommand{\rnf}{V_n}
\newcommand{\al}{\alpha}
\newcommand{\be}{\beta}
\newcommand{\ga}{\gamma}
\newcommand{\om}{\omega}
\newcommand{\Sch}{Schr\"{o}dinger \ }
\newcommand{\Schs}{Schr\"{o}dinger's \ }
\newcommand{\drm}{\mathrm{d}}
\newcommand{\und}{\underline}
\newcommand{\vs}{\vspace*{5mm}}
\begin{titlepage}
\title
{Unfinished History and Paradoxes of Quantum  Potential.\\
 I. Non-Relativistic Origin,  History  and Paradoxes}
 \vspace*{3mm}
 \author{ E. A. Tagirov\\
\small({Joint Institute for Nuclear Research, Dubna 141980, Russia,}
 \small {tagirov@theor.jinr.ru})}
\begin{abstract}
 This  is the  first of  two related papers analising and  explaining the origin,
 manifestations  and  parodoxical features   of the  quantum potential (QP) from the non-relativistic  and
 relativistic   point  of  view. QP arises in   the  quantum Hamiltonian,
  under various procedures of quantization of the natural systems, i.e.  the  Hamilton functions
  of which are the positive-definite quadratic  forms in  momenta with  coefficients   depending on the
   coordinates in ($n$-dimensional) configurational space $V_n$ endowed so  by a Riemannian structure.
   The result  of  quantization  may be  considered as quantum mecanics (QM) of a particle  in $V_n$ in  the  normal
   Gaussian system  of  reference  in  the globally-static  space-time $V_{1,n} $.
   Contradiction of  QP  to  the  Principles  of  General Covariance and Equivalence is  discussed.

    It is  found that actually   the  historically  first Hilbert
    space based     quantization  by E.~Schr\"{o}dinger (1926),  after revision in   the  modern framework of QM,
    also leads to  QP in the form that B.~DeWitt had  been found
      26 years later.   Efforts to  avoid   QP or reduce   its  drawbacks   are   discussed.
      The general conclusion is  that some form   of QP and a violation of the principles  of  general relativity
      which it  induces  are  inevitable  in the  non-relativistic   quantum  Hamiltonian.
       It is  shown also that  Feynman (path integration) quantization  of  natural systems  singles out
       two  versions of  QP, which
       both determine  two bi-scalar (indepedendent on choice of  coordinates)  propagators   fixing two
       different algorithms  of path integral calculation.

 In the  accompanying paper under the same  general title and  the  subtitle
 \emph{"The Relativistic  Point of  View"},   relation  of  the non-relativistic QP to
 the quantum  theory  of the  scalar field non-minimally coupled to the curved space-time metric is
 considered.

   \end{abstract}
   \end{titlepage}
\maketitle

 \section{Introduction}
 In the present  paper, the different  procedures of  the Hilbert space based quantization\footnote{This term is
used here
   in the   sense   considered in theoretical physics as construction of quantum mechanical counterpart
   to  a  class of  the  hamiltonian
  mechanical systems.  The more mathematically strict approaches  adopted  in  mathematical physics
  suggest that  quantization is  a map from  the   category  of commuting  objects to  another category
   where  the image-objects    are non-commutative.  Quantization, so  understood,  meets serious mathematical
   obstructions, see  \cite{Got1, Got2}.} of  the non-relativistic
 natural mechanical  systems will  be  analyzed  and  compared.
\emph{The (classical) natural systems} (the term originated  by E.~Whittaker and  re-animated
   by V.~N.~Arnold and  A.~B.~Givental \cite{arn}) are those whose  Hamilton functions  are
   non-uniform quadratic forms in   momenta $p_a$  with  coefficients $\om^{ab}(q)$ depending  on coordinates
 $ q^a, \quad a, b,\dots = 1,\dots ,n $ of configurational space $V_n $:
 \beq
    H^\mathrm{(nat)} (q, p; \om)  = \frac1{2m} \om^{ab}(q) p_a p_b +  V^\mathrm{(ext)}(q).
     \label{Hnat}
\nde
 It provides $V_n $  by a Riemannian structure
\beq
 \drm s_\mathrm{(\om)}^2 = \omega_{ab}(q)\drm q^a \drm q^b .\label{s2}
 \nde
 (Henceforth, subscripts $\mathrm{(\om)}$ and $\mathrm{(g)}$ will  denote objects related to  metric
 tensors $\om_{ab}$  and $g_{\al\be}, \quad  \al, \be, \dots = 0, 1,\dots ,n $  of $V_n$ and of $n+1$--dimensional
 space-time $V_{1,n}$ respectively)
 Thus, $H^\mathrm{(nat)}$ determines the dynamics  of  a natural  system   as     a  motion  of a
 point-like  particle  in  $V_n$,  to which a potential $V^\mathrm{(ext)}(q)$   acts  in  addition. So, the actual
    motion    of  a neutral   point-like  particle  in  the  external
  gravitation  (including  description of the  motion in  curvilinear coordinates  without   gravitation)
  treated  general-relativistically  as a curved space is a representative  case  of  a natural system.

  However,  there   is  a principal  alternative way to  construct  the non-relativistic QM of this
  simplest physical system  coupled to the geometrized  gravitation.
  Namely, to extract it  from  the   the
  general-relativistic quantum  theory of   scalar field  as the
  non-relativistic ($c^{-1}=0$) asymptotic  of  its  one-particle sector.  The one-particle subspace
  in  the particle-interpretable    Fock representation  of the canonically quantized field    can  be  defined (in  the asymptotical
  sense) even   when the  metric is  time-dependent,
   i.e.  $\om^{ab}= \om^{ab}(t,q)$.  This  approach will be  considered
   in the companion  paper \cite{Tag} under the  same title and subtitle \emph{"The Field-Theoretic  Point of
   View"} and compared with  conclusions  of  the present paper.  It should be noted  at once  that  \emph{these  two
   approaches  lead,  in  general, to  QM's which  do  not  coincide  completely}. This is one  of interesting
    results of  the  work  as  a whole.

  Despite that this   basic problem may be considered  as of
   little "practical" interest for  physics,   there is  an important  aspect of it.    The theory,   which is more
 general and  geometrically transparent than the standard  QM in  a potential  field, can serve as an  instrument
 for a deeper   insight on foundations of QM.  E.~\Sch  was  guided  just by such an  idea when he had proposed a
  method  of construction of   a  quantum  Hamiltonian  for  the  generic  natural    system  in the
  third \cite{Sch1} of  his  five   papers \cite{Sch2} of 1926,  by which  the  wave  mechanics had been founded.
   Apparently,  it  was the  first attempt  of  quantization in  the  sense which is  close
   to the  modern  meaning  of  the term in  theoretical physics. While this  step, which had not received a deserving
    attention, \Sch did not even  mentioned gravitation  or general  relativity at all.

    In  the present paper, we shall, on contrary,  analyze  whether  the two basic  general-relativistic
    principles - the Principle  of  Equivalence (referred further as PE)   and  the  Principle of  General Covariance,
    which     hold on the  classical level for any natural
    system,     are satisfied in a sense in the  corresponding QM.  It is a paradox that  the  both   of  the
    principles does not hold in the ordinary sense in  QM constructed by quantization of the natural systems,
     but satisfy them in  a restricted  sense in QM extracted from the general-relativistic quantum  theory of  scalar
     field.

   All  various procedures of quantization of Hamiltonian  systems with finite degrees  of freedom
    are ambiguous  or problematic to  be  mathematically rigorous. Therefore,  it seems more correct to speak
 on  a paradigm of  quantization  rather   than  on  an  well-established theory. See, e.g.,
 a  discussion of the  topic  by M.~J.~Gotay in \cite{Got1}.   As concerns  the level of  mathematical
 rigor of the present discussion,  the  best way    to  characterize
   it is  the  following amusing citation taken from  \cite{Stern}:
  \begin{quote} \textit{"...as Sir Michael Atiyah said in his closing lecture of the 2000 International
Congress in Mathematical Physics,..., Mathematics and Physics are two communities separated by a common
language."} \end{quote}
 Then, the  present work is from  the side of the  Physics community.
  We  will  be  mainly concerned   with  the  so called the Hilbert-space based  canonical quantization, which  is
  meant here as  a map
  \beqa
    q^a \longrightarrow \hat q^a ,  \ p_b \longrightarrow \hat p_b, && \mbox{so  that}\
    \{q^a,\, p_b\} = [\hat q^a,\,\hat p_b],     \label{qp}\\
        H^\mathrm{(nat)} \longrightarrow  \hat H, \label{HH}
  \ndea
  where  all "hatted" objects are   assumed  to have  representation as differential operators in
   the  Hilbert space $L^2 (\rnf; {\mathbb C};  \omega^{1/2} d^n q)$ and $\{.,\, .\}$ is the  Poisson bracket.
              The quantum Hamiltonian    $\hat H $   is assumed to  be constructed
   of the  "primary" quantum observables $\hat q^a,\,\hat p_b$   by some substantiated way. In  the  standard  canonical
     quantization, it is found by the
   straightforward  substitution (\ref{qp})into $H^\mathrm{(nat)}(q, p) $ and   some Hermitizing ordering.

        Specifically,  conclusions of
    analysis   of the following    approaches to  QM  of the natural systems
    by  the  present author  will  be  exposed  below descriptively or, at least, noted:
    \begin{itemize}
 \item
  \emph{\Schs variational approach}  (Section 2);
 \item
 \emph{revision of \Schs variational approach} by  the present author  (Section 3);
 \item 
   \emph{canonical quantization (\Sch-DeWitt ordering) }
    \cite{DW1}  and its generalization by  the  present  author (Section 4);
   \item 
  \emph{quasi-classical quantization} by  B.~S.~DeWitt \cite{DW2} (Section 5);
 \item 
  the  Blattner--Costant--Sternberg formalism in \emph{the geometric quantization} \cite{Snia, Kal} (Section 6);
 \item 
  \emph{path  integration, or the Feynman quantization} \cite{Fe, DW2, Mcl,  Chen, Dow, DOT,  Klein, Tag1}  (Section 7).
  \end{itemize}
  Some intermediate  conclusions from this  first  part  of  the  analysis  is  given  in  Section 8 and further
 they   will be compared
  in the  accompanying paper \cite{Tag}   with  the  asymptotic in $c^{-1} \rightarrow 0 $ of  the quantum theory of
     scalar field in  the  general globally static Riemannian
    space-time $V_{1,n}$  and  the proper frame of reference in   which  the  metric form of $V_{1,n}$  is
    \cite{Tag2, Tag1}:
    \beq
    \drm s_{(g)}^2 = g_{\al\be}(x) \drm x^\al \drm x^\be = c^2 \drm t^2 - \om_{ab}(q) \drm q^a \drm q^b , \quad
    \al,\be, \dots = 0, 1,\dots n ; \label{ds}, \quad  \{ct, q^a\} \sim {x^\al} \in V_{1,n}.
    \nde
 \section{Variational quantization of natural systems by \\ \Sch}
\Sch \cite{Sch1} searched a wave theory which plays the same role  \textit{w.r.t.} the Hamilton mechanics, that
     the Wave Theory  of  Light does \textit{w.r.t.} the Geometrical Optics.
  In  \cite{Sch1}, entitled  \textit{"On relation of  the Heisenberg--Born--Jordan quantum
mechanics to the  one of  mine"},  the third  of the seminal papers \cite{Sch2},  he constructed  a wave (quantum)
counterpart for the natural Hamilton function     $H^\mathrm{(\om)} (q, p)$
 as an extremal of the following functional (\Sch considered $\om_{ab} \equiv  \om_{ab} (q)$):
 \beq
   J^\mathrm{(Sch)}\{\psi\} = \int_{V_n} \omega^{\frac12}\,\drm^n q \,\left\{\frac{\hbar^2}{2m}
    \left(\frac{\ptl\psi}{\ptl q^a}\, \omega^{ab}\, \frac{\ptl\psi}{\ptl q^b}\right)
    + \psi^2 V^\mathrm{(ext)}(q)\right\} \label{Phi}
    \nde
  with the  additional condition
  $$
   \int_{V_n} \omega^{\frac12} \drm^n q \,\psi^2 = 1.
  $$
 It is important to note that \Sch considered  here \emph{the  real  wave functions} $\psi (q)$.
       Variation of  $J^\mathrm{(Sch)}\{\psi\}$ results in an  equation  for  eigenavalues $E$ of
         a differential operator in  the  space of functions $\psi (q)$, which  may be called  the  quantum  Hamiltonian :
  \beqa
      \hat H^\mathrm{(Sch)}\psi &=& E \psi, \label{Scheq} \\
 \hat H^\mathrm{(Sch)} &\defst & - \frac{\h^2}{2m}\Delta_\mathrm{(\om)} + V^\mathrm{(ext)}, \label{Hsch}
 \ndea
   It looks  as satisfying  to  the  conditions  which  are  implied   by GR: it  is generally covariant,
   i.e. a scalar \textit{w.r.t.} point  transformations   $q^a \rightarrow \tilde q^a (q)$
  and  satisfies to PE    which sounds  in  the  formulation by S.~Weinberg \cite{Wein}
  as follows:
\begin{quote}\textit{"... at every space-time point in an arbitrary
gravitational field it is  possible to choose "a locally inertial coordinate system", such that within a
sufficient small region of the point of question, laws of nature take the same form as
 in  an unaccelerated Cartesian coordinate system."}
\end{quote}
 According to eq.(\ref{ds}),  $\om_{ab}$ may  be,   in  particular,
   a  relic  of a general-relativistically  treated  gravitation and,  in this  sense,  PE can  be applied
   to $\hat H^\mathrm{(Sch)}$. A more fine question is: are  the  \Sch equation (\ref{Scheq})
   and  its time-dependent and,  further, general-relativistic  generalizations  are  "laws of nature"
      which must satisfy PE?  We shall return  to  it  in Sec.9. and in the companion paper \cite{Tag}

 It should be emphasized also that \emph{\Sch himself by  no  means  related  his quantization of the  natural
 systems  to gravitation or GR.} He considered it as an  instrument to  investigate the quantization  procedure
 itself  by application  it  to   mathematically more general classical cases than the simple potential ones.
 Just this  is our main aim  but  for  a more wide variety of quantization procedures and  in relation with GR.

\section{Revision of \Sch approach in  framework of  modern quantum  mechanics}
 In  the modern QM ,  $\psi(q)$ are actually complex functions from
  a  pre-Hilbertian  space $L^2 (\rnf; {\mathbb C}; \omega^{1/2} \drm^n q)$   with the  scalar  product
    \beq
     (\psi_1, \psi_2) \defst \int_{V_n} \ov\psi_1 \psi_2 \, \omega^{\frac12} \drm^n q , \qquad
     \psi \in  L^2 (\rnf; {\mathbb C};  \omega^{1/2} \drm^n q). \label{scpr}
     \nde
      The  physical sense of  \Schs functional (\ref{Phi})  is the  mean value of
     the  energy of  the system in  the  state $\psi(q)$.
    Instead, today we should  take the matrix elements of  energy:
       \beq   J^\mathrm{(modern)} \{\psi_1,\psi_2]\} = \int_{V_n} \left\{\frac1{2m}
    \overline{\hat p_a\psi_1}\, \om^{ab}(\hat q)\, \hat p_b \psi_2
    + V_{\mathrm{(ext)}}(q) \overline\psi_1 \psi_2 \right\}\om^{1/2} \drm^n q \label{psi2}
        \nde
    where $\hat q^a \defst q^a \cdot \hat {\mathbf 1}$ are
     the operators of  coordinates in the configurational space $V_n$ and
    $\hat p_a$  are the operators of  momentum canonically conjugate  to $\hat q^a$ :
    \beq
        [\hat q^a, \hat p_b] = i\h \delta^a_b . \label{pq}
    \nde
       They should be Hermitean(!) \textit{w.r.t.} the scalar product $(\psi_1, \psi_2)$.
       Hermitean momentum  operators for $V_n$ were introduced  first  by W.Pauli \cite{Paul} in 1933:
     \beq
               \hat p_a \defst  -i\h\om^{-1/4} \frac{\ptl}{\ptl q^a} \cdot \om^{1/4}, \label{Pau}
     \nde
     where "cdot" denotes the  operator product.
    Then,  substitution of this expression into $J^\mathrm{(modern)} \{\psi,\ \psi\} $
    and \Schs  variational  procedure give   the  eigenvalue  equation similar to eq.(\ref{Scheq})
    but with a different quantum hamiltonian $\hat H^\mathrm{(DW)}$
  \beq
  \hat H^\mathrm{(DW)}\psi \defst \hat H^\mathrm{(Sch)}\psi +  V^\mathrm{(qm)}(q)\psi =  E \psi,\quad
  V^\mathrm{(qm)}(q)\defst   - \frac{\hbar^2}{2m}\ \om^{-\frac14}\ptl_a (\om^{ab}\ptl_b \om^{\frac14})
  \label{HDW}
  \nde
   The   term $V^\mathrm{(qm)}$ was discovered for the first time   by DeWitt \cite{DW1} in  1952  in a  different
   formalism  of quantization, who called it \emph{the quantum  potential}, see Section 3 below. Surprising is
    that it \emph{depends on  choice of coordinates} $q^a$  (i.e., is  not a scalar \emph{w.r.t.} transformations
    of $q^a$ ).
    Also, it \emph{violates PE} if eq.(\ref{HDW}) is taken as a quantum "law of  Nature" for a particle
    in the gravitational field described by $\om_{ab}$ since
               \beq
          V^\mathrm{(qm)}(y) = - \frac{\h^2}{2m} \cdot \frac16 R_{\mathrm(\om)}(y) + O(y). \label{Vqm}
               \nde
   in the quasi-Cartesian (normal Riemannian) coordinates $y^a$  with  the  origin  at the
   point $q$ under consideration.

    Thus,  the  dogmas of GR and  QM are in  conflict here! Moreover, non-covariance  of  the  quantum
    potential implies  that  \emph{the  energy spectrum} and, after transition to  time-dependent  version of  the
    \Sch equation, \emph{the  dynamics depend on  choice  of  coordinates} in QM so  constructed!
   A heretical thought comes  here. \textit{Perhaps,   it was  a  good  fortune for the early stage
  of  QM  that \Sch did not realize the conflict: one may suppose, he  and his successors in development of
  the  wave mechanics (see \cite{Jam}, Sections 5.3, 6.1) would be  embarrassed to
 proceed!}

  Returning to expression of   $ \hat H^\mathrm{(DW)}$   with  account  of  eqs. (\ref{Hsch}) and (\ref{Vqm})
we see that  the zero-order term in the  quasi-Cartesian coordinates $y^a$ having been taken separately  is
"value" of a scalar object  in  these  particular  coordinates by its  geometrical  sense . However, in the full
Hamiltonian
  $ \hat H^\mathrm{(DW)}$, they are not scalars because  the non-invariance of  the  residual term  tangles
  the  situation in other coordinates. We shall call the such  terms  \emph{quasi-scalars} in  the  theory under
  consideration.

 \section{Discovery of quantum  potential by DeWitt and  \\
  generalization of  his approach}
  26 years after \Schs result, B.~S.~DeWitt \cite{DW1} had come to the hamiltonian $\hat H^\mathrm{(DW)}$
    by a procedure which may  be  called  the canonical quantization; it is a map:
    \beq
       q_a \rightarrow \hat q_a,\ p^a \rightarrow \hat p^a \ \Rightarrow \
  H^\mathrm{(nat)} (q, p) \rightarrow  H^\mathrm{(DW)} (\hat q, \hat p) \defst
  \frac1{2m} \hat p_a \om^{ab}(\hat q)\hat p_b +
    V^\mathrm{(ext)}(\hat q).  \label{nat1}
    \nde
  Here, the von Neumann rule
  \beq
       \hat f(q^1,\dots,q^n) = f (\hat q^1,\dots,\hat q^n) \label{Neu}
        \nde
  for definition  of the operator corresponding  to  a function  of classical observables,
   the  Poisson  brackets of  which vanish,
is  applied  for  definition $\hat \om^{ab} (q)$.

  As a differential  operator in $ L^2 (\rnf; {\mathbb C}; \omega^{1/2} \drm^n q)$,
\beq
\hat H^\mathrm{(DW)} (\hat q,  \hat p) = \hat H^\mathrm{(DW)} \defst \hat H^\mathrm{(Sch)} +
 V^\mathrm{(qm)}(q) \  \mbox{(!)}.\label{HDW2}
\nde
  Thus, \emph{the  revised version  of the \Sch  quantization coincides with   DeWitt's canonical
quantization!}
  Evidently, DeWitt himself   did not know the  original  \Sch  work \cite{Sch1}.

    DeWitt's result is related to  the particular  ordering of non-commuting  operators
  $\om^{ab}(\hat q),\hat p_a$.  Other (Hermitean) orderings
  (Weyl, Rivier,  \textit{et all}) are well known. Then, on the our level of quantization,
      \emph{why  not  to consider   Hermitean linear combinations  of  different orderings?}
   The simplest   class  of Hamiltonians so obtained form an  one-parametric family:
  \beqa
  \hat H^\mathrm{(\nu)}
  &=&\frac{2-\nu}{8m} \om^{ij} (\hat q) \hat p_i \hat p_j + \frac\nu{4m} \hat p_i  \om^{ij} (\hat q) \hat p_j +
  \frac{2-\nu}{8m}\hat p_i \hat p_j \om^{ij} (\hat q)  \nonumber\\
  &= & \hat H^\mathrm{(Sch)} +  V^\mathrm{(qm;\nu)}(q) \cdot \hat{\mathbf 1} \label{Hnu}\\
  V^\mathrm{(qm;\nu)}(q)&\equiv& V^\mathrm{(qm)}(q) + \frac{\h^2(\nu-2)}{8m}\ptl_a\ptl_b \om^{ab}.  \label{Vnu}
         \ndea
 DeWitt's  ordering corresponds  to $\nu = 2$. In  the  quasi-Cartesian coordinates $y^a$
 \beq
    V^\mathrm{(qm;\nu)} =  - \frac{\h^2}{2m} \cdot \frac{\nu}{12} R_{\mathrm(\om)}(y) + O(y). \label{Vnuy}
 \nde
 Thus, there  is  an  ordering  with $ \nu =0 $  for which the zeroth-order short distance term
 vanishes,   but  the non-zero residual term retains; it   means that PE is  satisfied  in  the  QM  if one considers
     $\hat H^\mathrm{(\nu=0)}$ (because there is  no curvature term at the point  of the  particle  localization)
   but  it is still not  covariant. Besides, it will be seen  in \cite{Tag} that $\nu=0$ does not agree with
   the requirements of  PE  to  the  relativistic  propagator.

  \section{Quasi-classical quantization}
  DeWitt   did not  take  notice  of the  non-invariance of  $V^\mathrm{(qm;\nu)}$,  referring to
 possibility to  transform it from  one  coordinate system  to  another, which is, of course,  not  invariance.
   However, evidently  he  had not been satisfied  by the result of the  canonical quantization.
   In 1957,  DeWitt \cite{DW2}    determined  quantum  Hamiltonian   as a differential operator
   in   $L^2 (\rnf; {\mathbb C}; \omega^{1/2} \drm^n q)$
  through  construction  of  quasi-classical  propagator ${\Bbb G}(q'',\, t''| q',\, t')$:
  \beq
 \psi(q'',\,t'')
 = \int_{V_n} \om^\frac12(q') \drm^n q' \,{\Bbb G}(q'',\, t''| q',\, t')\, \psi(q',\, t'), \label{G}
 \nde
 by generalization of the  Pauli  construction  for a charge in  e.m. field \cite{Pau} to the  case of  natural systems:
 \beqa &&{\Bbb G} (q'',\, t''| q',\, t') = \nonumber\\
 &&\qquad =\omega^{-1/4} (q'')\, D^{1/2} (q'',\, t'' |q',\, t' )\,\omega^{-1/4} (q')
 \, \exp\left(-\frac i\hbar  S(q'',\, t''|q',\, t') \right), \label{GDW}\\
 &&D (q'',\, t'' |q',\, t' )\defst \det\left(- \frac{\ptl^2 S}{ \ptl {q''}^i \ptl {q'}^j}\right)\quad
  \mbox{(the  Van Vleck  determinant)} \nonumber
\ndea
  and  $S(q'',\, t''|q',\, t')$ is  a  solution  of  the  Hamilton-Jacobi equation  for
 $H^{\mathrm (nat)} (q,p)$.
  Using the Hamilton-Jacobi equation DeWitt had found that, in a small neighborhood
  of  space-time point  $\{q', t'\}$,  the  propagator ${\Bbb G}(q"\,t"| q',\,t')$
  \textit{"nearly satisfies the \Sch equation"}.
   (Henceforth, $V^\mathrm{(ext)}\equiv 0$ is taken  for  simplicity.)
   \beq
    \frac{\ptl}{\ptl t''} {\Bbb G} (q",\,t"| q',\, t')=
   - \frac{\h^2}{2m} \Delta_{\mathrm (\om)} {\Bbb G}(q",\,t"|q',\,t')
    +\widetilde {V}^{\mathrm{(qm)}}(q'',\,t''){\Bbb G}(q",\,t"| q',\,t'). \label{G1}
      \nde
  where
 \beqa
    \widetilde{V}^{\mathrm{(qm)}}(q",\,t"|q',\,t')&\defst&  \frac{\h^2}{2m} f^{-1}(q",\,t"|q',\,t')
   \Delta_{\mathrm{\om}} f(q",\,t"|q',t') \nonumber \\
         &=& \frac{\h^2}{2m}\, \frac16 R_{\mathrm{(\om)}}(q", t")+ o(q"-q')+ o(t"-t'), \label{Vqmt}\\
    f(q",\,t"|q',t')&\defst& \om^{-\frac14}(q",\,t")\, D^{\frac12}((q",\,t"|q',\,t') \om^{-\frac14}(q',\,t').
    \nonumber
 \ndea
   So we see that $\widetilde{V}^{\mathrm{(qm)}}$    \emph{looks as a  scalar and  yet as violating PE}.
          Actually, $\widetilde{V}^{\mathrm{(qm)}}(q",\,t"|q',\,t')$ \emph{is a bi-scalar}  and
  thus  depends  on  choice of the  line
  connecting points $q'$  and $q"$.  If the geodesic lines  are chosen, then, in the
   asymptotic $q"\rightarrow q'$,  it is equivalent  to fixation of $q^a$  as the quasi-Cartesian
   coordinates $y^a$ and, thus, the  non-invariance of  the  quantum potential remains.
    \section{Geometric quantization of  natural systems}
  Geometric quantization is  oriented to  consider   $V_n$   with   non-trivial topologies, see e.g.
   the monograph by J.~Sniatycki \cite{Snia} and the paper \cite{Kal}. In the  latter paper, expansion by
    $c^{-2}$   of the Hamilton   function for the  the relativistic particle in the proper system  of  reference:
  \beq
  H^\mathrm{(rel)} (q, p) = mc^2 \sqrt{\hat{\mathbf 1} + \frac{2 H_\mathrm{(\om)}^\mathrm{(nat)} (q, p)}{mc^2}} \label{Hrel}
   \nde
    is  considered using the  Blattner--Kostant--Sternberg formalism.

   The zero-order  quantum potential is
   $V^{\mathrm{(qm)}}(q) = \frac{\h^2}{2m}\,\frac16 R_{\mathrm{(\om)}}(q)$, that is a scalar
   but the  geometric  quantization  is  a locally asymptotic theory by construction,
   and,  thus, merely supports DeWitt's and revised \Schs   local asymptotic quantum  potential.
            This paper is  interesting  also in  that  the  second  order  term in  the  asymptotic expansion
   in $c^{-2}$,     which is  quartic  in  the momenta, is  considered.
   The corresponding   potential is a rather
      complicate  scalar expression including  derivatives  and  quadratic expressions  of  the  curvature
      tensor. Thus, $\widehat{{H^\mathrm{(nat)}}^2} \neq \{\hat H^\mathrm{(nat)}\}^2$ and consequently, the  von
      Neumann rule does  not  work for  the polynomials of $H^\mathrm{(nat)}$.  An interesting problem  to study.

\section{Feynman quantization of natural systems}
  There  are  many papers devoted to  construction of the quantum propagator for
  natural systems by  path  integration  so  that the short-time  asymptotic of the propagator  would
  reproduce \Schs  original (invariant)   Hamiltonian,   
  However, it requires  some deformation
  of the  classic Lagrangean  with which the  path  integration starts usually, see, e.g., \cite{Chen, Dow} and, as a method of quantization
   is equivalent, on my opinion, to mere postulation  of  the \Sch original Hamiltonian. Instead,
   I shall  return to  the  original  idea of Feynman on path  integration \cite{Fe}, but with use results  of
   the  generalized  canonical quantization (Section 4 above) and  admit,  if necessary, QP in the quantum Hamiltonian generating
   the  original form of the  Feynman propagator. Fixation of QP is,   in  fact, quantization
   (in the  sense  accepted here)     of the  natural
   mechanics  under  consideration.

   The Feynman propagator  ${\Bbb G}^{\mathrm (F)} (q,\, t| q_0,\, t_0)$  \emph{is constructed}
 by division of  finite time interval  $t-t_0$  by $N \rightarrow \infty$  intervals
    $[t_I,\ t_{I+1}],\ I = 0,1,\dots, N-1, \ t_N = t$   of duration  $\epsilon = (t-t_0)/N$  as follows:
      \beq
  {\Bbb G}^{\mathrm(F)} (q,\,t| q_0,\, t_0) \defst
  \lim_{\substack{N\rightarrow\infty \\ \epsilon\rightarrow 0}}\
  \frac1{A^{N+1}} \int \exp\left(\int_{t_0}^t L_\mathrm {(eff)} \drm t \right) \prod_{I=1}^{N}
  \om_I^\frac12 \drm q_I , \label{gfe}
     \nde
   where $ A\defst (2\pi i\h\epsilon)^{\frac12 n}, \, q_I \defst q(t_I), \, \om_I \defst \om(q_I)$.
    A question  arises at once what is   the effective  Lagrangean   $L_\mathrm{(eff)}$?

    For  the  natural  systems Feynman's   choice would be
     \beqa
  L_\mathrm {(eff)}&=& L_\mathrm{(classic)} \nonumber\\
    &=& \frac{m}2 \om_{ab}(q) \dot{q}^a\dot{q}^b, \qquad \dot{q}^a \defst
  \frac{\drm q^a}{\drm t},
  \label{fe}
   \ndea
   i.e. the Lagrangean of  geodesic motion in  $V_n$ (the case of $V(q)\equiv 0$ is taken  for simplicity and
   straightforward comparison with  the relativistic field theory in \cite{Tag2}).
     Then each integration on interval  $ [t_I,\ t_{I+1}]$  is  taken along a geodesic connecting  $q_I$ and $q_{I+1}$ .
  However,  to have  the desired \Schs result
 \beq
       i\h \frac{\drm}{\drm t}\psi(q,t) = \hat H^\mathrm {(Sch)} \psi(q,t),
   \nde
  according to \cite{DW2, Dow}  et al., the  choice should be
 \beq
 L_\mathrm {(eff)} \equiv L_\mathrm{(classic)}
 - \ \frac{\h^2}{2m}\, \frac1{12} R_\mathrm{(\om)}(q) \label{eff}
  \nde
       to compensate the quantum potential term.
       But then the virtual classical motion between $q_I$  and $q_{I+1}$  will be not geodesical.
   Also, other  ambiguities arise  in  the  process of  calculation  of  a Hamilton  operator
   from the  path  integral (\ref{gfe}). Instead of reviewing  them,  further I expose   briefly  main points
   of a  special approach   the  initial  idea of  which is taken from paper by  D'Olivo and Torres \cite{DOT}
   but essentially modified in  \cite{Tag1} and consists of the following steps:

    {\textbf 1.} Consider the Hamiltonian representation of ${\Bbb G}_\mathrm{(F)} (q,\,t| q_0,\, t_0) $
     as a fold of the short-time propagators  in  the  configuration  space  representation \cite{Fe}:
     \beq
  {\Bbb G}^\mathrm{(F)} (q,\,t| q_0,\,t_0) \defst
    \lim_{N\rightarrow\infty} \int \, \prod_{K=1}^{N-1} \omega^{\frac12} (q_K)\, \drm^n q_K \prod_{J=1}^{N}
  <q_J|e^{-\frac\im\h \epsilon\hat H^\mathrm{(eff)} (\hat q, \hat p)}|q_{J-1}> \label{geff}
    \nde
 where $q_K =q(t_K)$.\\

 {\textbf 2.} It is natural  to suggest that  the effective Hamiltonian
 $\hat H^\mathrm{(eff)} (\hat q, \hat p)$ as a differential operator in   $L^2 (\rnf; {\mathbb C}; \omega^{1/2} \drm^n q)$
  is known  up to some effective potential $V^\mathrm{(eff)}(q)$, i.e.
 \beq
 \hat H^\mathrm{(eff)}  =  - \frac{\h^2}{2m} \Delta_{\mathrm (\om)} + V^\mathrm{(eff)}(q) \label{Veff}
 \nde
 Thus,  our task is  to  find  $ V^\mathrm{(eff)}(q) $ which  provides the  hamiltonian  form of propagator (\ref{geff})
 with  the  Lagrangean form (\ref{gfe})   so  that  $L_\mathrm{(eff)} \equiv L_\mathrm{(classic)}$    .

 {\textbf 3.}  To  calculate the  matrix elements in  configuration representation,  one  should to  express
       the  differential operator $-\h^2/(2m)  \Delta_{\mathrm (\om)}$  in  eq.(\ref{Veff}) through operators
       $\hat q^a, \ \hat p_b$ remaining $V^\mathrm{(eff)}$ still  undetermined. The  expression  depends  on the  rule
       of  ordering of $\hat q^a, \ \hat p_b $and $\om^{ab} (\hat  q)$    We take the one-parametric family of
       linear combinations  of different Hermitean  orderings introduced  in Section 4 :
        \beq
        \hat H^\mathrm{(eff)} (\hat q, \hat p) = \hat H^\mathrm{(\nu)}(\hat q, \hat p)-  V^\mathrm{(qm;\nu)}(\hat q)
        + V^\mathrm{(eff)}(\hat q).  \label{HV}
        \nde

  {\textbf 4.} Calculation  of the matrix elements within the  terms  linear in $\ep$
 using our generalized rule of ordering gives:
 \beqa
 {\Bbb G}^{(\nu)} (q'', t''|q', t') &=& \lim_{N\rightarrow\infty}
 \int \ \left(\frac{1}{2\pi \im\h\epsilon}\right)^{\pi N/2}
 \prod_{I=1}^{N-1} \sqrt{\om (q_I)}\ \drm^n q_I \nonumber\\
 &\times&\prod_{J=1}^{N-1} \frac{\left(\tilde{\sqrt\om}\right)^{(\nu)} (q_{J-1},\,q_J)}{[\om(q_J)
 \om(q_{J-1})]^{1/4}} \exp\left\{\frac{\im}{\h} \epsilon {\tilde L}^{(\nu)}_\mathrm {(eff)}\left(q_{J-1}, q_J;
 \frac{\Delta q_J}{\epsilon}\right)\right\}, \label{K3}\\
 \Delta q_J &\equiv&  \{\Delta q^i_J \defst q^i_J - q^i_{J-1}\}.\nonumber
 \ndea
  Here   $(\tilde  {\sqrt\om})^{(\nu)} (q_{J-1}, q_J), {\tilde L}^{(\nu)}_\mathrm{eff}\left(q_{J-1}, q_J, \Delta q_J/\epsilon
 \right)$ are the  kernels  of  the  corresponding  operators in configurational representation.
  They  are expressed, respectively, through functions ${\sqrt\om (q)} $ and
   \beq
 L_\mathrm{(eff)}^{(\nu)}\left(q, \frac{\Delta q_J}{\epsilon} \right)
 \defst L_\mathrm{(classic)} \left(q,\ \Delta q_J/\epsilon \right) -
 V^\mathrm{(eff)}(q)  + V^\mathrm {(qm;\nu)}
 \nde
 along the  rule:
 \beq
 {\tilde f}^{(\nu)}\left(q_{J-1}, q_J\right) = \nu
f(\bar q_J) + \frac{1-\nu}2 \left(f(q_{J-1}) + f(q_J) \right), \quad \bar q_J \defst \frac 12 (q_J + q_{J-1})
\label{tilde}
 \nde
 which follows  from  the  general rules  of  quantization  of  Beresin and Shubin,  \cite{BSh}, Chapter 5, in
 terms  of  the  kernels of operators.\\

{\textbf 5.}
 Then, the product enumerated by $J$  in eq.(\ref{K3}) should be  represented  as
a product of exponentials of  the  values   of  the   classical action on the  intervals $[q_{J-1},\, q_J]$, that
is as a product of factors  of the form
\beq
\exp\left\{\frac{\im}{\h} \epsilon
 L'_\mathrm{(eff)}\left(q'_J,\ \Delta q_J/\epsilon\right)\right\},
\label{L'} \nde where, in the exponent, the value of some effective Lagrangian $ L'_\mathrm{(eff)}(q, \dot q) $
(in general, it differs from $L_\mathrm{(eff)}^{(\nu)}$ )   stands,  which is taken at a point $ q'_J \in
[q_{J-1},\, q_J]$
 chosen so that  to represent the  exponent  as $L_\mathrm{(classic)}$ . To obtain  the representation,
all functions of $q_{J-1},\ q_J, \bar q_J$  under the product in $J$ should be expanded into the Tailor series
near the  point $q'_J$ up to terms quadratic in  $\Delta q_J$, since only such terms contribute to the integral in
eq.(\ref{K3}). Further, one should include the contribution of the pre-exponential factor to the exponent in a
form of an additional QP.\\

{\textbf 6.} The problem  of evaluation of  the  integrands  is  divided into the two principally different
choices of  the  point $ q'_J \in [q_{J-1},\, q_J]$:
\begin{itemize}
 \item \emph{Case A)}. The  end point evaluation of  the integrands:  $ q' =  q_{J-1}$ or
$ q' =  q_J$ i.e., $ q'$  is taken at the ends of the segment $[ q_{J-1},\,  q_J]$. However, to avoid appearance
of terms proportional to $\dot q_J $  and, thus,  of  parity  violation in  the resulting
 $L'_\mathrm {(eff)}(q, \dot q) $ and also  for agreement with rule (\ref{tilde}),
 the  quantum image   $\tilde f^{(\nu)}( q_{J-1},\,   q_J) $  of  the generic function $f(q)$
 should depend on the endpoints symmetrically .
 In the necessary  approximation, it is
 \beqa
 \tilde f^{(\nu)}( q_{J-1},\,  q_J) &=&  \frac 12 f( q_{J-1}) +  \frac 12 f( q_J) + \frac{\nu}8 \bigl(\ptl_i f( q_{J-1})
- \ptl_i f( q_J)\bigr) \Delta q_J^i \nonumber\\
&+ &  \frac{\nu}{16} \bigl(\ptl_i \ptl_j f( q_{J-1}) + \ptl_i \ptl_j f( q_J)\bigr) \Delta q_J^i\Delta q_J^j + O
\left((\Delta q_J)^3\right). \label{fb}
\ndea
 \item \emph{Case B)}.  The intermediate point evaluation of the  integrands:
$ q'_J = (1-\mu)  q_{J-1}  + \mu  q_J, \quad 0< \mu < 1 $,  i.e.,  $ q'_J \in ( q_{J-1},\,  q_J)$. For the generic
function $f(q)$ on the interval $( q_{J-1},\,  q_J)$   one has
 \beq
 {\tilde f}^{(\nu)}\left( q_{J-1},
 q_J \right) = f( q'_J)  + (\frac 12 - \mu)\ptl_i f( q'_J) \Delta  q_J^i + \frac 12 (\frac{2-\nu}4 -\mu + \mu^2
)\ptl_i\ptl_j f( q'_J) \Delta q_J^i\Delta q_J^j, \label{fmu} \nde
\end{itemize}

 Referring to more  details  of rather complicate calculation in \cite{Tag1}, now  only  final conclusions will be
 given, which are
 important for further discussion. In  the  both  cases,  we  come again  to noninvariant quantum potentials
  $V^\mathrm{(eff)}(q)$ which  we will denote  $V^\mathrm{(eff;A)}(q)$ and  $V^\mathrm{(eff;B)}(q)$.\\

 In  the  case A),  transformation  of  the  Hamiltonian  form (\ref{geff})  of  the  Feynman propagator
 ${\Bbb G}^\mathrm{(F)}$ to its  Lagrangean form  (\ref{gfe}) with  $L_\mathrm{(eff)} \equiv L_\mathrm{(classic)}$
 is  possible only  if $\nu = 2 $  and  then the   generally non-invariant
 potential  $V^\mathrm{(eff)} (q)$ has the form
 \beq
   V_\mathrm{A}^\mathrm{(eff)} (q)  = -  \frac{\h^2}{12 m} (2 \om^{ij}\om^{kl}
+\om^{ik}\om^{jl})\ptl_i\ptl_j \om_{kl} - \frac{\h^2}{16 m}(2 \ptl_i\om^{ij}\ptl_j \ln\om + \om^{ij} \ptl_i
\ln\om\,  \ptl_j \ln\om) \label{VA} ,
 \nde
 and, in  the  quasi-Cartesian coordinates $y^a$, it is
\beq
 V^\mathrm{(qm)}(y)   = - \frac{\h^2}{2m} \cdot \frac16 R_\mathrm{(\om)}(y) + O(y). \label{VAR}
 \nde
 That is, in the case A) the quantization map  $ H_0 \rightarrow \hat H_0^\mathrm{(eff; A)}$, which
 may  be  called   \emph{the Feynman  quantization},   coincides remarkably with  the result of
 the    revised \Sch and  "canonical"  DeWitt quantizations in the zeroth order of   local asymptotic. In the
 complete form,  it  differs  from  the latter,  what is   worth  of a further study.

In the  case B) the  same  approach leads to  (the result does not depend on the  parameter $\mu$)
  \beqa
   V_\mathrm{B}^\mathrm{(eff)} (q)
  &=& \frac{\h^2}{4m} \left( \frac{\nu + 2}{4} \om^{km}\om^{ln}\om^{ij} -
\frac{\nu - 2}{4} \om^{im}\om^{jn}\om^{kl}- (\nu - 2) \om^{im}\om^{kn} \om^{jl} \right)
\ptl_i\om_{mn}\ptl_j\om_{kl},\label{VB}\\
 V_\mathrm{B}^\mathrm{(eff)} (y) &=&  - \frac{\h^2}{2m} \, \frac13 R_{(\om)} (y) + O (y^2). \label{VBR}
  \ndea

 Asymptotic  local  expressions (\ref{VAR}) and  (\ref{VBR}) for  quantum potentials $V_{\mathrm A}^\mathrm{(eff)}$  and
  $V_{\mathrm B}^\mathrm{(eff)}$ (but  not  the  complete potentials (\ref{VA}) and  (\ref{VB})) follow also
  from the  study of the Pauli-DeWitt and Feynman quantizations by G.~Vilkovysskii \cite{Vil}. He
  worked  in  the framework of  the  formalism  of  proper  time  in  relativistic quantum mechanics
  $V_{1,n}$ (so  that   his  expressions for  potentials include $R \equiv R_{g}$, the scalar  curvature of $V_{1,n}$).
   From his  argumentation one  may  conclude that  just the case A) is the  preferred  one.

 However, if the set of points  $ q_J $  is  considered  as a lattice in $V_n$, then the case {\textbf A)} corresponds to
  evaluation  of the  integrand  as the arithmetic mean of its values  on the  adjacent nodes of  the  lattice.
     The case {\textbf B)} corresponds  to  evaluation at the  mean point of  the edges. Thus, these  two cases
      fix  two  different ways  of lattice calculation  of  the  path  integral (\ref{gfe}) with  same  integrand
      $ L_\mathrm{(eff)}= L_\mathrm{(classic)}$,  that  give,  in principle,  different  propagators.

More important  is  that the  complete  expressions  for  QPs (\ref{VA}) and (\ref{VB}) are defined for any
coordinates $q$ whereas the asymptotic expressions  (\ref{VAR}) and (\ref{VBR}) are suited only for
quasi-Cartesian coordinates according to  argumentation of  Section 4.  But the most  impotant, though
paradoxical,  conclusion is  that \emph{QPs  in the both  cases are  not scalars}  of the general transformation
$\tilde q^a = \tilde q^a(q) $ Thus,   we encountered  again  with the non-invariance of quantum Hamiltonian
despite that the   Lagrangean  form  (\ref{gfe}) of  the propagator is invariant.

\section{Intermediate  conclusion}

Firstly, we see a deep conflict in  QM between the requirements  of observability (hermiticity) and   invariance
with  respect to transformations $\tilde q^a =  \tilde q^a (q)$ (general covariance).
   Apparently, the  non-invariance of QM seems to be a general property
of the standard quantization approaches based on a Hilbert space of states.
It looks very strange,  but conceptually it can be explained  by that the quantum operators of observables and,
particularly, of coordinates $\hat q^a = q^a \cdot \hat {\mathbf 1} $, imply some concrete classical measurments
over the  quantum natural system which is  an  open  system   and depend on  choice of them, while the classical
$q^a$ are considered as any of abstract arithmetizations of space points. Further, this conception  probably is
related to the ideas that  information on the open  quantum system always includes more or less information on the
apparatus which provides an  information on the  system. Discussion of  these ideas may be  found  in  papers by
M.~Mensky \cite{Men1}  and C.~Rovelli \cite{Rov}.
 However,    our  consideration perhaps leads  further: not only
information  on  the  state of  object but  also  information  on  its  dynamics  contains a mixture  of
information  on  the  measuring device (of the particle's position,  in  our  case) through the additional  terms
in  the Hamilton  operators.  This  thought  is   supported  by  that  \emph{QP is  not  a specifity  of curvature
of the  space only and    related  to  choice of curvilinear  coordinate  systems in  the  flat   space , too.}

  Of course, this  issue needs a deep  analysis and  still I can give  nebulous speculations  encouraged by the
following words of  Leon Rosenfeld concerning QM:
\begin{quote}\textit{"... inclusion of specifications of
conditions  of observation into description of phenomena is not  an arbitrary decision but a necessity dictated by
the  laws themselves of evolution of  the phenomena and mechanism of observation them, which makes of   these
conditions an integral part of  the decsription  of  the phenomena"}.\footnote{Requotation   from the Preface
\emph{"To Soviet readers"}  for the Russian edition of the book by I.Pigojine and I. Stengers \emph{Order out of
Chaos.Man's new dialogue with nature}, "Progress" Publishing House, Moscow, 1986. (inverse translation from
Russian  of the present author (E.T.))}\end{quote}

 From this  point  of view,    propagator
 ${\Bbb G}^\mathrm {(\nu)} (q'', t''|q', t') $,  eq.(\ref{K3}), is the amplitude of  transition  of   the particle
 from point $q'$
 to point $q''$, the  position  of  which  is  subjected to  continual observation  by means  of local coordinates
 $y^a$  on each  infifnitesimal section  of  possible  trajectory.  To this end, it is sufficient to  take
 the (quasi-)scalar  terms of
 the quasi-classical approximation in  the  Hamiltonian  and therefore the  amplitude  is  a bi-scalar.
 (Apparently, this paradox  is  an  analog of  the one known as \emph{"the continually observed  kettle boils never"},
 see, e.g. \cite{Sud}.)So, the local  quasi-Cartesian coordinates $y^a$   add   no  information except the  scalar curvature.

  On  contrary,    any version of  the full  Hamiltonian
 $\hat H$ is  used  to   prepare a  state $\psi(q)$  with  use of curvilinear  coordinates $q^a$. Their coordinate  lines
 are  determined  by  $n$  curvatures   which  are equal to zero only  in the  case of  the  Cartesian coordinates
 which exist  only  in  $E_n$. They  affect as forces on motion  of the  particle, see, e.g., \cite{Syn}, Chapter 1 ,
 which are different  for  different choice  of the  coordinates $q^a$.

     The difference between the versions of quantum Hamiltonian originally formulated  in the Cartesian coordinates and
thereupon transformed to the the spherical ones and the one which is immediately  formulated  in  the  latter
coordinates had been noted by B.Podolsky  in 1928 \cite{Pod}. The  way  out of the problem,  which he had
proposed, is a particular case of the following more general postulate:
   \beq
      \hat H^\mathrm{(Pod)} \defst  \hat\om^{-\frac14}\hat p_a \om^{\frac12}\hat \om^{ab} \hat p_b \hat\om^{\frac14}
      \equiv \hat H^\mathrm{(Sch)}= -\frac{\h^2}{2m} \Delta_{(\om)} \label{pod}
   \nde
However, this is equivalent  to the direct  postulation  of  the  desired  invariant result while \Sch wanted to
find the quantum  Hamiltonian   from  a general variational  principle.   Further, why one may not to dispose any
appropriate degrees of  $\hat\om(q)$ between multipliers  of $\hat H^\mathrm {(DW)}$ or even  take normalized
 Hermitean linear combinations of such  disposals? There is  a continuum  of  such  generalizations of $\hat H^\mathrm {(Pod)}$
 with zero and as well as non-zero QPs. An answer may  be that $\hat H^\mathrm{(Pod)}$ and
apparently the  the  mentioned  disposals of degrees of $\hat\om$ with  the resulting zero QPs are discriminated
by their invariance with  respect to  transformations  of  coordinates. Nevertheless, their multiplicity causes
some  dissatisfaction and needs  of  further  study. Besides, as it will be seen in the companion paper
\cite{Tag}, all these versions of the  theory do not  satisfy  to the Principle of Equivalence from   the
general-relativistic  point  of  view.

 Another way  to invariant quantum   Hamiltonian for
the natural  systems, which is based on  use of non-holonomic coordinates,  have been proposed by H.~Kleinert in
his monumental
 monograph \cite{Klein} on  path integration and by M.~Mensky \cite{Men2}.  However,  this interesting non-metric
  approach    will   be out yet  of  the scope of the  present  paper  where we hold  only the    metric  approaches.

Summing up,     the considered  or just now mentioned approaches to  quantization of the natural systems
discriminate   three  preferred  classes of the  quantum Hamiltonians     which are  characterized  by  the values
of the constant   $\xi$   in the  quasi-scalar  term  of  the  local  asymptotics:
 \beq
        V^\mathrm{(qm)}(y)  = - \frac{\h^2}{2m}\, \xi R_{(\om)}(y), \label{xi}.
 \nde
  These  values  and  fomalisms which  fix  them  are:
  \begin{itemize}
   \item[$\xi = \frac16 $] $\longleftarrow$ \{ canonical  and quasi-classical quantization  by  DeWitt; revised \Sch
   variational  approach and    Feynman   quantization in
   present author's version  along the evaluation  rule on the lattice for functions in integrands:
   \beq
   q^\prime \in [q_J, q_{J+1}],  \quad  f(q^\prime)  =  \frac{f(q_{J+1})+ f(q_J)}2 ; \label{fjj}
    \nde
   \item[$\xi = \frac13 $] $\longleftarrow$ \{quasi-classical quantization by Vilkovysky,  Feynman quantization
    present author's version  along the evaluation  rule on the lattice for functions in integrands:
   \beq
   q^\prime \in (q_J, q_{J+1}),  \quad  f(q^\prime)  =   f(\frac{q_{J+1}+ q_J}2); \label{qjj}
    \nde
   \item[$\xi =0 $] $\longleftarrow$ \{\Schs original approach (no QP), Rivier ordering in canonical quantization
   (there is QP  but  the  quasi-scalar  term  vanishes), the  Podolski  postulate and  its  generalizations,
   quantization in non-holonomic coordinates \cite{Klein}\}.
    \end{itemize}
 There are  very  interesting  problems to  study  related  to  each of the three versions  of  quantization.
 For example, it would be important to  extend them for   the  algebra   of  polynomials  of  momenta $p$
 with coefficients  depending  on canonically conjugate   $q$. This
 would  transfer  the  theoretic-physical  conception  of  quantization, which  is adopted in  the  present paper
 and  the  most  of references therein,     to  a more  mathematically  refined   level.

  It should not be forgotten  also that  the (revised) \Sch variational approach to quantization can  apparently
  be of interest
 for  application  to  topologically non-trivial cases of  $V_n$,  while,  in   the  present work, the  triviality
 is  intrinsically implied.

 However, still, we shall follow  the more
 pragmatic  way and  consider   the  possible  values  of  $\xi$ from a  point  of  view  of  general-relativistic quantum
    theory  of   the  linear  scalar field,  of  which the  nonrelativstic asymptotic ($c^{-1} \rightarrow 0$ )  of the one-particle sector
    of  which    should  produce  quantum  mechanics of  the  natural  systems. In  particular, it
     will  be shown there that
    just the theory with  $\xi = 1/6 $ is  in  accord with  the  Principle  of  Equivalence as it  formulated by
    S.~Weinberg, see Sec.3 and  supported  by the conformal symmetry if $n=3$.

\section{Acknowledgement}
Thanks to  Professors P.~Fiziev, V.~V.~Nesterenko and  S.~M.~Eliseev  for useful discussions ànd consultations.

\end{document}